\documentclass[final,5p]{elsarticle}
\usepackage{jasr}
\usepackage{framed,multirow}
\usepackage{amssymb}
\usepackage{natbib}
\usepackage{latexsym}
\usepackage[normalem]{ulem}
\usepackage{url}
\usepackage{amsmath}
\usepackage{subcaption}
\usepackage{xcolor}
\definecolor{newcolor}{rgb}{.8,.349,.1}
\usepackage[citebordercolor=white]{hyperref}
\usepackage[left]{lineno}

\journal{Advances in Space Research}

\begin{document}

\begin{frontmatter}

\title{Accretion Flow Properties of MAXI J1910-057/Swift J1910.2-0546 During Its 2012--13 Outburst} 

\author[1]{Sujoy Kumar \snm{Nath}}
\author[2,1]{Dipak \snm{Debnath}\corref{cor1}}
\cortext[cor1]{Corresponding author: 
  email: dipakcsp@gmail.com}
\author[1,2]{Kaushik \snm{Chatterjee}}
\author[3]{Arghajit \snm{Jana}}
\author[4]{Debjit \snm{Chatterjee}}
\author[1]{Riya \snm{Bhowmick}}

\address[1]{Indian Centre For Space Physics, 43 Chalantika, Garia Station Road, Kolkata, 700084, India}
\address[2]{Institute of Astronomy Space and Earth Science, Sector II, Salt Lake, Kolkata 700091, India}
\address[3]{Institute of Astronomy, National Tsing Hua University, Hsinchu 30013, Taiwan}
\address[4]{Indian Institute of Astrophysics, Koramangala, Bengaluru, Karnataka 560034}

%\received{}
%\finalform{}
%\accepted{}
%\availableonline{}
%\communicated{}

\begin{abstract}

Galactic black hole candidate MAXI J1910-057/Swift J1910.2-0546 was simultaneously discovered by MAXI/GSC and Swift/BAT 
satellites during its first outburst in 2012. We study the detailed spectral and temporal properties of the source in a broad
energy range using archival data from Swift/XRT, MAXI/GSC, and Swift/BAT satellites/instruments. Low frequency quasi periodic 
oscillations are observed during the outburst. The combined 1-50 keV spectra are analyzed using the transonic flow solution 
based Two Component Advective Flow (TCAF) model. Based on the variations of soft and hard X-ray fluxes, their hardness ratios 
and the variations of the spectral model fitted parameters, we find that the source has evolved through six spectral states. 
We interpret this spectral state evolution to be a result of the release of the leftover matter from the pile-up radius 
due to a sudden rise of viscosity causing a rebrightening. We show a possible configuration of the evolution of accretion
flow during the outburst. From the spectral analysis with TCAF model, we estimate the probable mass of the black hole to 
lie in the range 6.31 $M_\odot$ to 13.65 $M_\odot$, and the source distance is estimated to be $1.9-8.3$~kpc from transition 
luminosity considerations.

\end{abstract}

\begin{keyword}
\KWD X-Rays:binaries \sep stars: individual: (MAXI J1910-057) \sep stars:black holes \sep accretion:accretion discs \sep radiation:dynamics \sep shock waves
\end{keyword}

\end{frontmatter}

\section{Introduction}

Stellar mass black holes residing in binaries can be detected by the electromagnetic radiations emitted from the matter accreting 
onto the black hole (BH) from its companion star. During the process of matter falling towards the black hole, a part of its 
gravitational potential energy gets converted into heat due to turbulent viscosity, which is then radiated in entire electromagnetic 
wavebands, i.e., from radio waves to $\gamma$-rays. High-energy X-rays come from regions near the BH event horizon, which carry 
important information for studying the properties of the BHs. 

Many of these black hole X-ray binaries (BHXRBs) spend most of the time in the quiescent state with very low X-ray luminosity 
($L_X \sim 10^{30-33}$ ergs/s; Tetarenko et al. 2016), and occasionally go into the outburst state when the X-ray luminosity 
of the source rises by several orders of magnitude ($L_X \sim 10^{37-38}$ ergs/s; Tanaka \& Shibazaki 1996). 
For low mass X-ray binaries, matter from the companion star accretes onto the black hole via Roche-lobe overflow and forms 
an inward-spiraling accretion disk due to its non-zero angular momentum. When the viscosity of the matter in the outer disk 
increases, the outburst is triggered and matter moves inward (Shakura \& Sunyaev 1973; Chakrabarti \& Titarchuk 1995; 
Ebisawa et al. 1996; Chakrabarti 2013). During an outburst, the temporal and spectral properties of BHXRBs change rapidly. 
In general, four different spectral states, namely, hard state (HS), hard-intermediate state (HIMS), soft-intermediate 
state (SIMS), and soft state (SS) are observed during an outburst of the transient BHXRBs based on the contribution of 
high energy (hard) and low energy (soft) X-rays in the total X-ray flux (Fender et al. 2004; Homan \& Belloni 2005; 
McClintock \& Remillard, 2006). The so-called `q'-diagram, i.e., hardness-intensity diagram (HID: Miyamoto et al., 1995; 
Homan et al., 2001; Belloni et al., 2005) or a more physical two component accretion rate ratio-intensity diagram 
(ARRID: Jana et al., 2016) shows the evolution of spectral properties with time. One can distinguish different spectral 
states from the different branches of the HID (Remillard \& McClintock, 2006). Generally, outbursting BHs show spectral 
states in the following sequence, HS (rising) $\rightarrow$ HIMS (rising) $\rightarrow$ SIMS (rising) $\rightarrow$ SS $\rightarrow$ 
SIMS (declining) $\rightarrow$ HIMS (declining) $\rightarrow$ HS (declining) (Belloni et al., 2005; Belloni 2010). 
Power-density spectra (PDS) of hard and intermediate spectral states sometimes show a peaked noise called 
quasi-periodic oscillations (QPOs) (Psaltis et al. 1999; Nowak 2000; Casella et al. 2004; Ingram \& Motta 2019 and references therein). 
While only a few BHXRBs show high-frequency QPOs (HFQPOs) with frequencies in the range 40-450~Hz, low-frequency QPOs (LFQPOs) in the 
range 0.1-30~Hz are rather common. Depending on their various properties (centroid frequency, Q-value, rms amplitude, and noise), 
these LFQPOs are classified into three types: A, B, and C (Casella et al., 2005). Monotonically evolving type-C QPOs are generally 
observed in the HS and HIMS, whereas type-A and type-B QPOs can be observed occasionally in SIMS (Remillard \& McClintock, 2006; 
Nandi et al. 2012; Debnath et al. 2013). No QPOs are observed in the SS.

The X-ray spectrum of a BHXRB can be interpreted phenomenologically as a combination of two components: a multicolor blackbody 
component of thermal soft photons along with a power-law type tail of non-thermal hard photons. The thermal blackbody component 
is believed to be emitted from an optically thick, geometrically thin Keplerian flow of the standard disk (Novikov \& Thorne, 1973; 
Shakura \& Sunyaev, 1973). Low-energy photons emitted from this standard disk get inverse Comptonized from a `Compton' cloud 
(Sunyaev \& Titarchuk, 1980, 1985)  containing hot electrons and generate the power-law tail of the spectrum. Various models 
in literature attempt to explain the nature of this Compton cloud, e.g., magnetic corona (Galeev et al., 1979), disk-corona 
model (Haardt \& Maraschi, 1993; Zdziarski et al., 1993), jet-base model (Kalemci et al., 2005; Markoff, Nowak, \& Wilms, 2005), 
two-component advective flow (TCAF) model (Chakrabarti \& Titarchuk 1995, hereafter CT95; Chakrabarti 1997). According to the 
TCAF model, the accretion disk consists of two components: an optically thick, geometrically thin, highly viscous Keplerian flow 
on the equatorial plane, and an optically thin, low viscous sub-Keplerian (low angular momentum) component encapsulating the 
Keplerian flow for a review of TCAF, see Chakrabarti 2018). At a certain distance from the BH, the centrifugal pressure dominates 
over gravitational attraction, and the supersonic sub-Keplerian matter gets slowed down to become subsonic through a shock 
transition and gets piled up behind the centrifugal barrier to form the CENtrifugal pressure supported BOundary Layer (CENBOL). 
This CENBOL, which is a repository of high-energy electrons, acts as the Compton cloud and emits the hard component of the spectra. 

The TCAF model has been implemented into HEASARC's spectral analysis software package {\fontfamily{pcr}\selectfont XSPEC} 
(Arnaud, 1996) as a local additive table model (Debnath et al. 2014, 2015) after generating model {\it fits} file using large 
number of theoretical model spectra. To fit a spectrum, TCAF model requires four independent physical flow parameters, namely 
the accretion rates of the Keplerian and the sub-Keplerian components, location of the shock (i.e., size of the Compton cloud) 
and compression ratio, i.e., ratio of the post-shock to pre-shock matter densities. Additionally, a mass parameter, giving 
the mass of the BH, and one normalization factor (that gives the ratio of emitted to observed X-ray spectrum) are also needed. 
One can obtain the best fitted values of these physical parameters which are directly related to the accretion process and 
infer on the accretion dynamics, as has been done for several BH candidates (Debnath et al. 2017; Mondal et al. 2016; Shang et al. 2019; 
Chatterjee et al. 2021b). Intrinsic source parameters, such as the mass of the BH (Molla et al. 2016; Chatterjee et al. 2021a) 
and jet/outflow X-ray flux (Jana et al. 2017; Debnath et al. 2021) also has been estimated from spectral analysis with the TCAF model.

Galactic transient BHXRB MAXI J1910-057 was discovered simultaneously by MAXI/GSC (Usui et al. 2012) and Swift/BAT (Krimm et al. 2012) 
on 2012 May 30-31 (MJD 56077-78). The first Swift Target of Opportunity (ToO) observation performed on 2012 June 1 localized the 
source at (R.A., Dec)(J2000) = ($19^{h}10^{m}22.78^{s},-05^\circ47'58.0''$) within a $3.5''$ radius of uncertainty (Kennea et al. 2012a).
Its optical/IR counterpart was observed on 2012 June 1, which suggested a BH binary nature (Rau et al. 2012; Kennea et al. 2012a). 
Optical photometry of the source on 2012 July 9 showed short timescale variations with a period of $\sim 2.2$~hr, indicating a short 
orbital period (Lloyd et al. 2012). However, on 2012 August 29 no such variations could be found, and optical spectroscopy suggested 
a longer period of $>$ 6.2~hr (Casares et al. 2012). Initially the source was in the soft state with a dominating thermal disk 
component (Kennea et al. 2012b; Kimura et al. 2012). A soft to hard state transition was reported on 2012 July 25 (Nakahira et al. 2012),
and a steady jet emission in radio was detected on 2012 August 3 (King et al. 2012) as usually seen in the hard states of BHXRB 
outbursts (Remillard \& McClintock, 2006). On 2012 August 22, the source was observed to be in the hard state by INTEGRAL 
(Bodaghee et al. 2012). Analysis of the XMM-Newton observations of the source showed the presence of reflection features, which 
indicates the possibility of a truncated inner accretion disk or a retrograde spinning BH (Reis et al. 2013). Analysis of the 
multi-wavelength lightcurves revealed a flux dip in all wavebands which might be related to a mass-transfer instability and 
the time delay of the occurrence of the dip between different wavebands might represent the viscous timescale of the disk 
(Degenaar et al. 2014). Nakahira et al. (2014) estimated the mass of the BH to be in the range 2.9 $M_\odot$ to 12.9 $M_\odot$.
Recently, the source has again been found to be in an outburst by MAXI/GSC on 2022 February 4 (Tominaga et al. 2022).
The source was also detected in radio wavelengths by Arcminute Microkelvin Imager Large Array (Williams et al. 2022), in optical 
wavelengths with the MITSuME 50 cm telescope Akeno (Hosokawa et al. 2022) and the Zwicky Transient Facility (Kong et al. 2022). 
By 2022 March 20, the source was observed to be already in quiescence with the Las Cumbres Observatory (Saikia et al. 2022).

In this paper, we study the spectral and temporal properties of MAXI J1910-057 during its 2012-13 outburst using Swift/XRT, 
Swift/BAT, and MAXI/GSC data with the TCAF model to investigate the physical properties of the accretion flow and to have 
a better estimate of the mass and distance of the source. In \S2 we briefly discuss the observation, data reduction, and 
subsequent analysis procedures. The analysis results are presented in \S3. Finally, in \S4 we summarize our results and 
draw the conclusions.

\section{Observation and Data Analysis}

We study the BH source MAXI J1910-057 during its 2012-13 outburst using Swift/XRT, Swift/BAT and MAXI/GSC data. The three instruments, 
when used simultaneously, cover an energy range of $\sim0.3-150$~keV. However, Swift/XRT data below 1 keV is ignored due to the presence 
of spectral residuals in some of the observations, and Swift/BAT data over 50 keV is ignored due to poor signal to noise ratio (S/N).
In our analysis, we use a total of 34 observations of the source obtained with the Swift/XRT (1-10 keV range) with target IDs 00032480, 
00524642, 00529076 and 00032521; Swift/BAT (15-50 keV range), 
and MAXI/GSC (7-20 keV range) between 2012 June 4 (MJD 56082) and 2012 November 14 (MJD 56245). In these 35 observations, there are 6 
observations during which only Swift/XRT, 28 observations during which Swift/XRT and MAXI/GSC, and 1 observation during which all 
the three instruments are used simultaneously. We use HEASARC's software package {\fontfamily{qcr}\selectfont HEASoft} version 6.26 
for our analysis.

The source was very bright in X-rays and Swift/XRT observations were heavily affected by pile-up (Kennea et al. 2012a). After the 
initial localization with the Photon Counting (PC) mode, Windowed Timing (WT) mode was used to obtain subsequent observations to 
reduce pile-up effects (Kennea et al. 2012b). Level 1 data files are processed with the task {\fontfamily{qcr}\selectfont XRT\_PIPELINE} 
to produce clean level 2 event files. To examine and eliminate the effects of pile-up, the method described by Romano et al. (2006) 
is followed. An annular region of outer radius $60''$ and suitable inner radius to eliminate the pile-up is selected in the 
{\fontfamily{qcr}\selectfont XSELECT} task to extract source spectra and light curves (with 0.01s time binning). Background spectra 
are obtained from the neighboring regions using a circle of radius $65''$. Using the tool {\fontfamily{qcr}\selectfont XRTMKARF} 
corresponding ARFs are created, and the appropriate RMFs are obtained from the {\fontfamily{qcr}\selectfont CALDB}. All spectra are 
rebinned to have at least 20 counts per bin with the tool {\fontfamily{qcr}\selectfont GRPPHA}. Since the source and background 
extraction regions are different, BACKSCAL keywords in the spectrum files are corrected using the tool 
{\fontfamily{qcr}\selectfont ahbackscal}.

Swift/BAT spectra are generated following the standard procedures. The {\fontfamily{qcr}\selectfont batbinevt} task is used to create 
detector plane image ($.dpi$) files. Detector masks are generated using the task {\fontfamily{qcr}\selectfont bathotpix}. To apply 
mask-weighting to the event files, {\fontfamily{qcr}\selectfont batmaskwtevt} task is used. A systematic error to account for the 
residuals in the response matrix is applied to the BAT spectra using the task {\fontfamily{qcr}\selectfont batphasyserr}. 
The {\fontfamily{qcr}\selectfont batupdatephakw} task is used to correct the ray-tracing keywords. Finally the response matrices 
are generated using the {\fontfamily{qcr}\selectfont batdrmgen} task. \href{http://maxi.riken.jp/mxondem/}{MAXI on-demand} process 
web tool is used to obtain the MAXI/GSC spectra (Matsuoka et al. 2009).

For spectral analysis, we use HEASARC's spectral analysis software package {\fontfamily{qcr}\selectfont XSPEC} version 12.10.1 
(Arnaud, 1996). We fit all the spectra with TCAF model based fits file as a local additive table model in 
{\fontfamily{qcr}\selectfont XSPEC} coupled with the absorption model {\fontfamily{qcr}\selectfont phabs}. 
To fit a BH spectrum with the TCAF model, one needs to supply four flow parameters: 
(\romannumeral 1) the Keplerian disk accretion rate ($\dot{m}_d$ in $\dot{M}_{Edd}$), 
(\romannumeral 2) the sub-Keplerian halo accretion rate ($\dot{m}_h$ in $\dot{M}_{Edd}$), 
(\romannumeral 3) the shock location ($X_s$ in Schwarzschild radius $r_s=2 GM_{BH}/c^2$), and 
(\romannumeral 4) the dimensionless shock compression ratio ($R = \rho_+/\rho_-$, ratio of the post-shock to the pre-shock 
matter density). Two additional input parameters, i.e., mass of the BH ($M_{BH}$ in $M_\odot$) and the model normalization ($N$) 
are also required. In our analysis, as the mass of the considered BH candidate is not previously determined from dynamical methods, 
initially we keep it as free while fitting spectra with the TCAF model. We obtain a best fitted value of $M_{BH}$ from each spectrum 
fit, and we refit all the spectra with $M_{BH}$ fixed to the average of these best fitted values. As there is presence of radio jets 
(King et al. 2012), the model normalization is also kept free, as it can vary greatly in presence of jets (Jana et al. 2017; 
Chatterjee et al. 2019). 

We use \href{http://maxi.riken.jp/star\_data/J1910-057/J1910-057.html}{MAXI/GSC} and \href{https://swift.gsfc.nasa.gov/results/transients/weak/SWIFTJ1910.2-0546/}{Swift/BAT} 
daily average lightcurve data obtained from their corresponding websites to study the long term variations of the source X-ray 
fluxes and the hardness ratios (HRs), i.e., ratios between fluxes of various energy bands (GSC 2-4 keV, GSC 4-10 keV, 
BAT 15-50 keV). We generate power density spectra (PDS) with the {\fontfamily{qcr}\selectfont powspec} task of the XRONOS package from 
the 0.01s time binned Swift/XRT lightcurves to search for low-frequency QPOs. The data of each observation is subdivided into several 
intervals each containing 4096 newbins and a PDS for each interval is generated which are normalized such that their integral gives 
the squared rms fractional variability and the expected white noise level is subtracted. These individual PDSs were averaged to 
generate a single PDS which was then geometrically rebinned with a series of step 1.07. 
The PDS were then modeled with broken power-law and Lorentzian models in {\fontfamily{qcr}\selectfont XSPEC}
to find the break frequencies and the QPO frequencies.

Note: In the TCAF model {\it fits} file (v0.3.2) used in the paper, we have incorporated some important changes in the model code. 
For the generation of the initial version (v0.1) of the TCAF model {\it fits} file, roughly original TCAF code 
(Chakrabarti \& Titarchuk 1995) was used (e.g., Debnath et al. 2014, 2015). In this version, for a better understanding of the 
reflection component by the hot photons from the CENBOL to the Keplerian disk, a logarithmic radial grid of the pre-shock Keplerian 
disk (starting from $X_s$) is used. A few more numerical corrections are also made. At present, the model {\it fits} file is 
publicly available on a demand basis. To get the latest version of the model {\it fits} file, we request to contact the 
corresponding author of this paper.

\vspace{0.5cm}
\section{Results}

\begin{figure*}
\vskip 0.5cm
  \centering
    \includegraphics[angle=0,width=14cm,keepaspectratio=true]{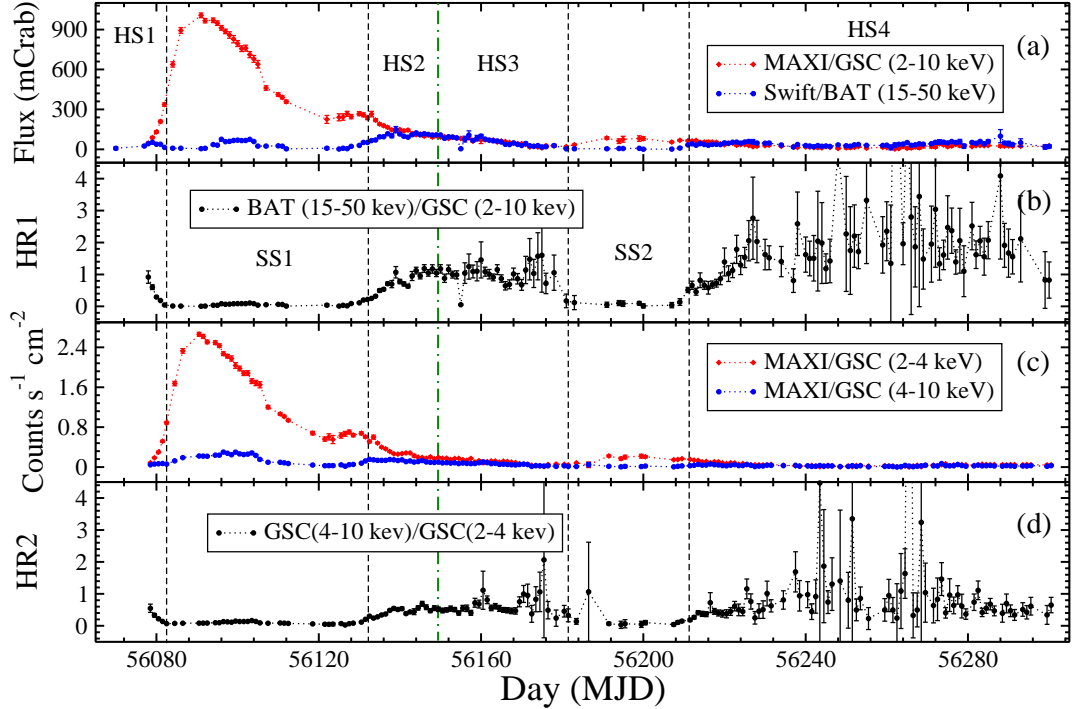}
	\caption{Variations of (a) 2-10 keV MAXI/GSC flux and 15-50 keV Swift/BAT flux in units of mCrab 
	                       (b) HR1, i.e., the ratio of 15-50 keV Swift/BAT flux to 2-10 keV MAXI/GSC flux 
			       (c) MAXI/GSC flux in 2-4 keV and 4-10 keV range in units of $Counts$ $s^{-1} cm^{-2}$  
			       (d) HR2, i.e., the ratio of 4-10 keV to 2-4 keV MAXI/GSC flux are shown with time (in MJD).
	         Vertical dashed lines mark the transition between spectral states. The vertical dot-dashed (green) line 
		 marks the transition between HS2 and HS3.}
\end{figure*}

MAXI J1910-057 is a transient X-ray binary which started its first outburst in 2012 May (Usui et al. 2012; Krimm et al. 2012). Swift and 
MAXI monitored the source for almost 8 months until 2013 January, after which the flux dropped below the lower limits of detection. 
However, radio (ATCA), UV (Swift/UVOT) and X-ray (Swift/XRT) observations in 2013 May indicated that the source was still in the hard 
state and the outburst was continuing (Tomsick et al., 2013). We analyzed observational data starting from 2012 June 4 (MJD 56082) 
when the source transitioned out of the rising hard state to 2012 November 14 (MJD 56245) when the source was already in the declining 
hard state. We report our analysis results in the following sub-sections.

\subsection{Temporal Properties}

We use MAXI/GSC and Swift/BAT on-demand/archival data to study lightcurve profiles, hardness-ratios, and Swift/XRT archival data 
for finding QPOs, after generating power-density spectra from the low time-binned lightcurves.

\subsubsection{Outburst Profile}

We plot different energy range lightcurves and their hardness ratios (HRs) in several 
panels of Fig. 1. In panel (a) we show the variation of 15-50 keV Swift/BAT flux (blue) and 2-10 keV MAXI/GSC flux (red), and 
their ratio is plotted in panel (b) as HR1. In panel (c), variation of MAXI/GSC flux in 4-10 keV (blue) and 2-4 keV (red)  
energy bands are shown, and their ratio is plotted in panel (d) as HR2. 

As it can be seen from Figure 1(a) and 1(c), the total GSC flux (2-10 keV) as well as the GSC soft flux (2-4 keV) increased rapidly, 
attained a maximum around 2012 June 12 (MJD 56090), and then gradually decreased. Fluxes in both of the energy ranges continued 
decreasing till 2012 September 8 (MJD 56178), and then increased slightly for a period of $\sim 33$ days.  %till MJD 56211 
After that, both of these decreased slowly till the end of our analysis period. 

GSC hard flux (4-10 keV) increased slowly and attained a maximum on 2012 June 18 (MJD 56096), 6 days later than the soft 
and total fluxes. Swift/BAT flux (15-50 keV) also increased slowly and attained a local maximum around this time. Both of these 
fluxes remained constant around this maximum value for $\sim 8$ days and then decreased slightly for the next $\sim 26$ days.
From 2012 July 22 (MJD 56130) the hard GSC flux increased briefly and then decreased to a roughly constant level, whereas 
the BAT flux increased to a level comparable to the total GSC flux (2-10 keV) during the period. It attained its maximum 
on 2012 July 31 (MJD 56139) and gradually decreased afterward. After 2012 September 8 (MJD 56178) the BAT flux declined 
below the GSC total flux and remained at the low level for $\sim 33$ days.  
After that, it again increased and remained roughly constant till the end of our analysis period.

\subsubsection{Hardness Ratio}

We show variation of the hardness ratios in Figs. 1(b) and 1(d). It is evident from the plots, starting from a high value at the 
onset of the outburst on 2012 May 31 (56078), both of the HRs (HR1 and HR2) decreased rapidly till 2012 June 4 (MJD 56082). 
The HRs remained almost constant at this low level for $\sim44$ days, after which they increased and became %till MJD 56127
roughly constant at a high value which is comparable to the first day of the outburst. After 2012 August 31 (MJD 56170) the HRs deviated 
from this constant level and became somewhat variable keeping a high value. They declined rapidly after 2012 September 8 (MJD 56178) 
and remained at a low value until 2012 October 9 (MJD 56209) after which both of the HRs increased to a high level till the end 
of our analysis period.

\begin{table*}[h]
%\small
 \addtolength{\tabcolsep}{-1.5pt}
 \centering
 \caption{QPO Properties}
 \label{tab:table1}
        \resizebox{1 \textwidth}{!}{
        \begin{tabular}{cccccccc}
\hline
	Obs Id &    UT Date   &  Day  & $\nu_{brk}^a$ & $\nu_{QPO}^b$ & $\Delta\nu^b$ & Q$^c$ &  rms \\
               & (yyyy-mm-dd) & (MJD) &    (Hz)       &    (Hz)       &    (Hz)       &       &  (\%) \\
           (1) &      (2)     &  (3)  &     (4)       &    (5)        &     (6)       &  (7)  &  (8) \\
\hline

		00032521004  &  2012-08-13  &  56152.5  & $ 0.81^{\pm0.15} $ & $ 2.60^{\pm0.11} $ & $ 0.46^{\pm0.19} $ &  5.68  &  8.65   \\
		00032521006  &  2012-08-19  &  56158.7  & $ 1.06^{\pm0.60} $ & $ 2.63^{\pm0.11} $ & $ 0.23^{\pm0.16} $ & 11.54  &  6.45   \\
		00032521010  &  2012-09-01  &  56171.3  & $ 0.82^{\pm0.27} $ & $ 2.80^{\pm0.20} $ & $ 0.38^{\pm0.12} $ &  7.97  &  8.97   \\
		00032521036  &  2012-10-27  &  56227.2  & $ 0.63^{\pm0.21} $ & $ 1.26^{\pm0.05} $ & $ 0.26^{\pm0.16} $ &  4.81  & 15.68   \\
		00032521044  &  2012-11-12  &  56243.0  & $ 0.09^{\pm0.07} $ & $ 1.52^{\pm0.07} $ & $ 0.21^{\pm0.17} $ &  7.20  & 20.45   \\

\hline
        \end{tabular}}
\vskip 0.2cm
\noindent{
	\leftline{$^a$ $\nu_{brk}$ is the observed break frequency as obtained from broken powerlaw fitting.}
       \leftline{$^b$ $\nu_{QPO}$ is the observed QPO frequency and $\Delta\nu$ is the full width at half maximum (FWHM), which are obtained by fitting}
       \leftline{the QPO bump on the PDS with a Lorentzian function.}
       \leftline{$^c$ Q = $\nu_{QPO}$/$\Delta\nu$ is the coherence factor indicating the sharpness of the QPO.}
}
\end{table*}

\subsubsection{Power Density Spectra (PDS)}

We study the fast Fourier transformed (FFT) power density spectra (PDS) generated from the Swift/XRT lightcurves of time 
resolution $0.01$~s for signatures of low frequency variabilities.
Among the 35 observations considered, QPOs were found 
in only 5 observations when the source was in the harder states (HS3 and HS4; see \S3.2.1). 
The band-limited noise (BLN) continuum is modeled with a broken power-law and QPO humps are modeled with a Lorentzian profile
to determine the centroid frequencies, Q-values, rms amplitudes, etc. The obtained results from the QPO fits are mentioned 
in Table 1. Among the five QPOs, three were observed in the third harder state (HS3) and the 
remaining two were observed in the fourth harder state (HS4).
Clear breaks in the low frequencies were seen in the PDSs and
the QPO frequencies were observed to increase in HS4 
(Figure 6(d)). We show the PDS observed on 2012 August 13 (MJD 56152) in Figure 2, where a QPO was seen at a frequency 
$2.60^{\pm0.11}$~Hz with Q-value 5.68 and 8.65\% rms variability. Based on the properties of the observed QPOs 
(centroid frequency, Q-value, rms), we categorize them as type-C QPOs, following the general criterion (Casella et al. 2005). 
The presence of clear breaks in the PDS and type-C nature of the QPOs indicate the source is in the harder states.

\begin{figure}[h]
%\vskip 1.5cm
%  \centering
    \includegraphics[angle=270,width=0.48\textwidth]{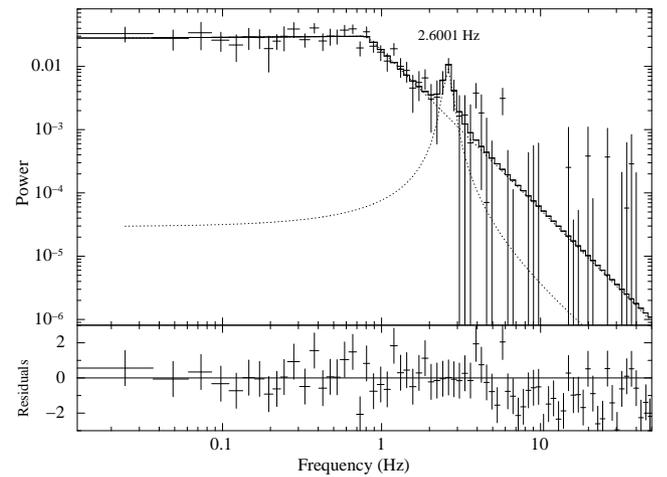}
	\caption{Continuum fit of the PDS obtained from the Swift/XRT observation on 2012 August 13 (ObsID = 00032521004) and the residuals. 
	A QPO of frequency 2.6 Hz can be seen.}
\end{figure}

We also investigate the correlation of QPO frequency and break frequency for MAXI J1910-057. The frequency of the
LFQPOs and the break frequencies of the broad-band noise components are known to correlate strongly in BHXRBs, atoll sources,
and millisecond X-ray pulsars (Wijnands \& van der Klis 1999; Belloni et al. 2002). In order to investigate the correlation,
we have used ADS' Dexter data extraction applet (Demleitner et al. 2001) to estimate the QPO and break frequencies of BHCs 
from Figure 2a of Wijnands \& van der Klis (1999). Then we have superimposed our observed QPO frequencies and the corresponding 
break frequencies of the five QPOs (red diamond points) along with the QPO obtained by Reis et al. (2013; blue circle point), 
and the results are shown in Figure 3. From the Figure, we can see that MAXI J1910-057 also follows a similar trend as other 
black hole candidates. We calculated the degree of correlation between the QPO frequency and the break frequency from the total 
sample of QPOs from Wijnands \& van der Klis (1999) and the QPOs of MAXI J1910-057. The Pearson correlation coefficient comes out 
to be $r(17)=0.96, p<0.00001$, and the Spearman's correlation coefficient comes out to be $\rho=0.94$, which signifies a strong 
positive correlation between QPO frequency and corresponding break frequency for BHXRBs.

\begin{figure}[h]
\vskip 0.8cm
%  \centering
    \includegraphics[angle=0,width=0.48\textwidth]{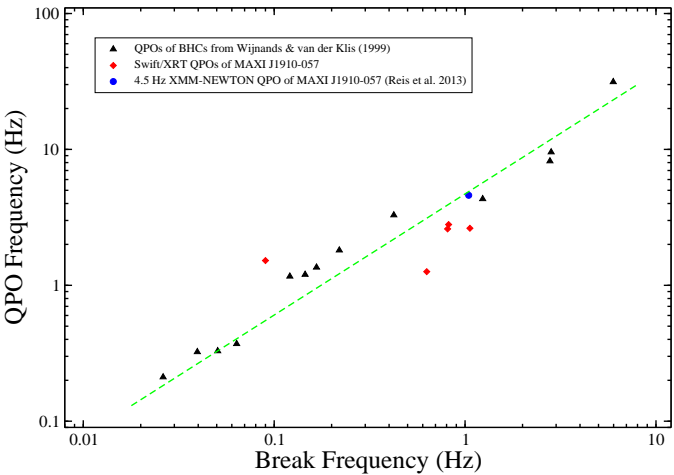}
	\caption{Correlation of break frequency vs QPO frequency for a collection of BHCs are shown with triangles (black), taken from the Figure 2b
	         of Wijnands \& van der Klis (1999). QPOs found in our study of MAXI J1910-057 are shown with diamonds (red). 
	         The circle (blue) shows the 4.5 Hz QPO found by Reis et al. (2013).}
\end{figure}

\subsection{Spectral Properties}

For studying the spectral properties of MAXI J1910-057 in a broad energy range, we use combined data from Swift/XRT, 
Swift/BAT and MAXI/GSC instruments. Nakahira et al (2014) used {\fontfamily{qcr}\selectfont TBabs$\times$(diskbb+Nthcomp)} 
model to determine various parameters, e.g. inner disk temperature, inner disk radius, photon index, etc. from this model 
and observed how these parameters evolve through various spectral states.
This phenomenological model, however, does not reveal the underlying accretion dynamics or the important radiation processes 
in the accretion disk. It models the hard component of the spectra as the thermally Comptonized part, but does not give us any
information about the location, size, geometry, etc. of the elusive `Compton cloud' where the Comptonization takes place.
To investigate the dynamics and the geometry of the accretion flow, we take a different approach and use the physical TCAF
model to analyze the spectra. We fit all the spectra with combined {\fontfamily{qcr}\selectfont phabs$\times$TCAF} model. 
We incorporate 3\% systematic error for the XRT spectra, as is advised in the XRT data extraction 
\href{https://www.swift.ac.uk/analysis/xrt/spectra.php}{webpage} for WT mode data of bright sources.
Initially, we fitted the spectra with the TCAF model keeping the BH mass free. From each spectral fit, we get the best-fitted value of
$M_{BH}$. This preliminary result is shown in Table A.
The $M_{BH}$ values were observed to vary between 7.4 $M_\odot$ and 13.5 $M_\odot$.
We take the average of these best-fitted mass values, which comes out to be 9.97 $M_\odot$. Then we kept $M_{BH}$
frozen at this value and refitted all the spectra to obtain the final result. The final results are shown in Table 2.
We show a TCAF model fitted spectrum and its residual in Figure 4, 
which was observed on 2012 June 23 (MJD 56101; Obs ID 00032480008) when the source was in SS1. 
The neutral hydrogen column density ($N_H$) was kept free.
The evolution of TCAF fitted total accretion rate ($\dot{m}_d+\dot{m}_h$), disk accretion rate 
($\dot{m}_d$), halo accretion rate ($\dot{m}_h$) and the accretion rate ratio (ARR = $\dot{m}_h/\dot{m}_d$) during the outburst 
are shown in panels (a), (b), (c) and (d) of Figure 5 respectively. Variation of TCAF model fitted 
mass ($M_{BH}$), shock location ($X_s$), shock compression ratio ($R$) and QPO frequency evolution are shown in panels (a), (b), (c), and (d) 
of Figure 6 respectively.

\subsubsection{Evolution of the Spectral states}

Unlike most BHXRBs, MAXI J1910-057 showed a different kind of spectral state evolution than the generally observed pattern. 
While the outburst profiles and the hardness ratios provided a basic estimate, our spectral analysis results
allowed us to clearly differentiate between the spectral states. Here we discuss the evolution of the spectral parameters,
e.g., the disk ($\dot{m}_d$) and halo ($\dot{m}_h$) accretion rates and their ratios as ARR, evolution of the shock location ($X_s$)
and strength ($R$) etc. through different spectral states.
It is to be noted that, in our previous studies of BHXRB outbursts with the TCAF model, it was seen that when an outbursting source transits 
between hard and soft state, the TCAF parameters doesn't evolve monotonically, but rather show some abrupt changes before entirely going to 
the soft/hard region (Chatterjee et al. 2020; Jana et al. 2020). This indicates that the source has transited to the intermediate states. 
However, in the 2012-13 outburst of MAXI J1910-057, no such prominent changes were seen during the transitions. 
Hence we define hard and hard-intermediate spectral states collectively as the harder state (HS), and soft and soft-intermediate spectral states
collectively as the softer state (SS).

\begin{figure}
  \centering
    \includegraphics[angle=270,width=0.48\textwidth]{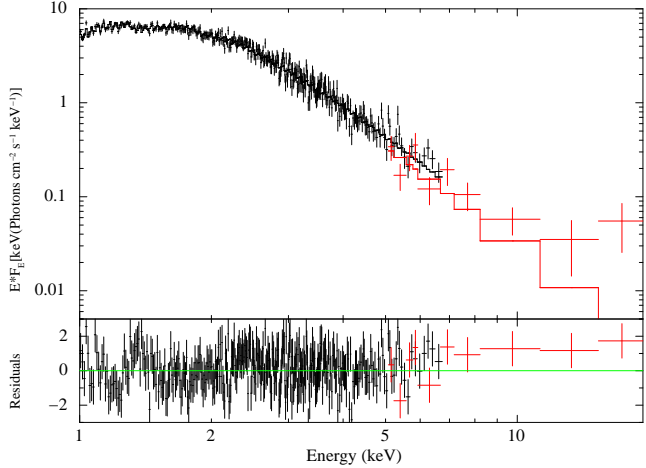}
    \caption{TCAF model fitted spectra on 2012 June 23 (ObsID = 00032480008). The black points show Swift/XRT data in the 1-7 keV band and the red points show MAXI/GSC data in the 5-20 keV band.} Model fitted spectral residuals are shown in the bottom panel.
\end{figure}

\begin{figure*}

  \centering
    \includegraphics[angle=0,width=14cm,keepaspectratio=true]{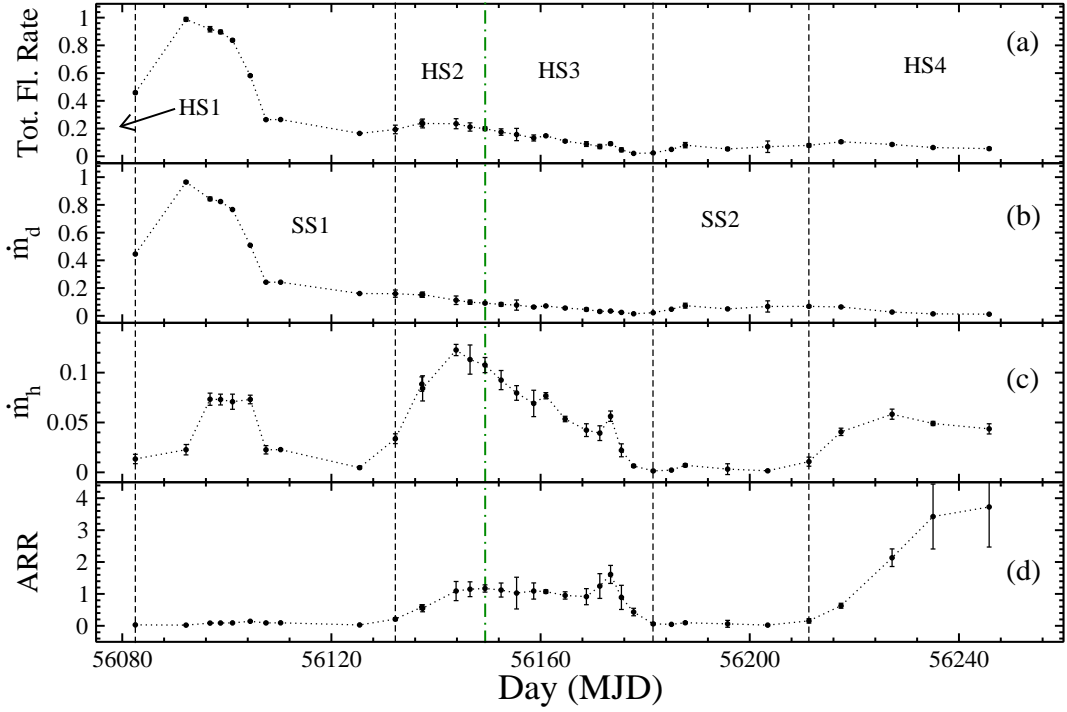}
	\caption{Variations of TCAF model fitted (a) total accretion rate ($\dot{m}_d$ + $\dot{m}_h$) in units of $\dot{M}_{Edd}$,
						 (b) Keplerian disk accretion rate ($\dot{m}_d$) in units of $\dot{M}_{Edd}$,
						 (c) sub-Keplerian halo accretion rate ($\dot{m}_h$) in units of $\dot{M}_{Edd}$,
					     and (d) the accretion rate ratio (ARR = $\dot{m}_h$/$\dot{m}_d$)
	        are shown with time (MJD). Vertical dashed lines mark the transition between spectral states. The first vertical dashed line marks
	        the start of the SS1 as no spectral analysis could be done in the HS1. The vertical dot-dashed (green) line 
		    marks the transition between HS2 and HS3.}
\end{figure*}

\textit{(\romannumeral 1) First Harder State (HS1):} At the start of the outburst, the HRs decreased rapidly from a high value 
(HR1=0.92, HR2=0.55) for a period of 4 days. We classified this period as the first harder state (HS1).
As there were no available spectra in this period of the outburst, we could not get any information from spectral analysis in this state of the outburst. 

\textit{(\romannumeral 2) First Softer State (SS1):} Our spectral observations start from 2012 June 4 (MJD 56082) when the source 
made a transition into the first softer state (SS1). The soft X-ray fluxes dominated the hard X-ray fluxes, and the HRs reduced to $\sim0.1$.
The disk accretion rates ($\dot{m}_d$) dominated the halo accretion rates 
($\dot{m}_h$) throughout this state resulting in a low ARR value. Initially, both of the accretion rates started to increase, 
and $\dot{m}_d$ attained a maximum on 2012 June 14 (MJD 56092). 
ARR was found to be in a range of low values ($\sim 0.03-0.14$) in this phase of the outburst as there was high dominance 
of the $\dot{m}_d$ over $\dot{m}_h$. A weak shock was observed which remained almost stationary close to the black hole 
($X_s\sim50r_s$). The source remained in this softer state till 2012 July 24 (MJD 56132). We defined 2012 July 24 (MJD 56132) 
as the SS1 to HS2 transition day, as we see a rise of the halo rate from this day onward.

\begin{figure*}
\vskip 1.5cm
  \centering
    \includegraphics[angle=0,width=14cm,keepaspectratio=true]{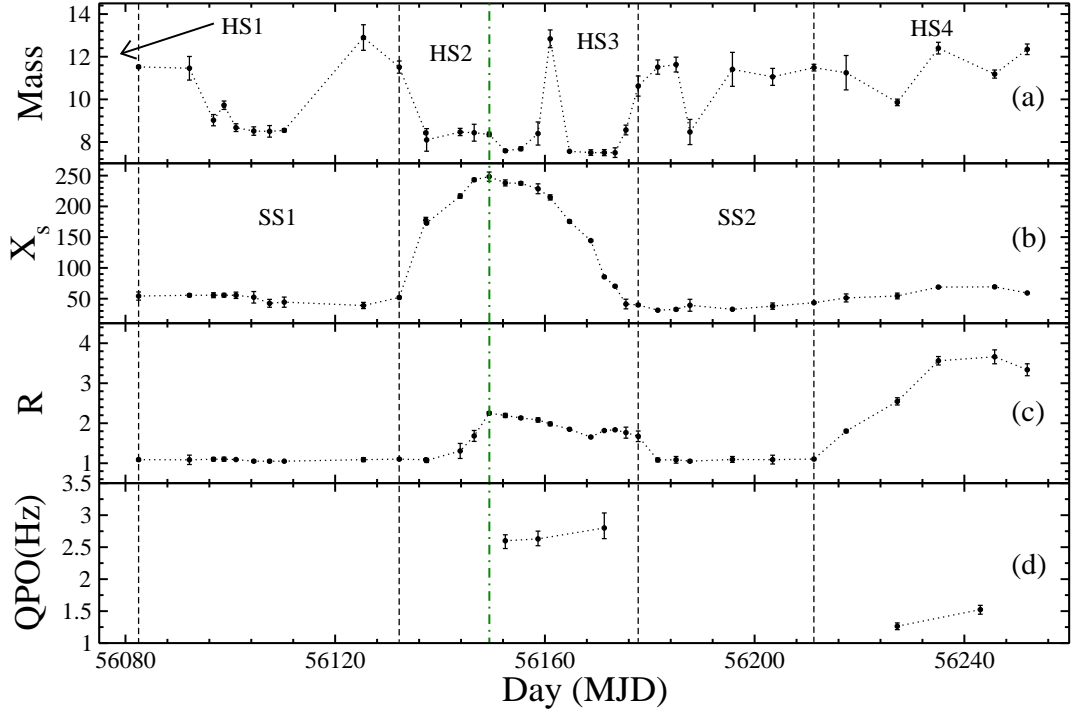}
	\caption{Variations of TCAF model fitted 
	                     (a) mass of the BH ($M_{BH}$) in units of $M_\odot$,
						 (b) the shock location ($X_s$) in units of Schwarzschild radius ($r_s$),
	                     (c) the shock compression ratio ($R$),
					 and (d) the evolution of QPO frequency 
	        are shown with time (MJD). Vertical dashed lines mark the transition between spectral states. The first vertical dashed line marks
	        the start of the SS1 as no spectral analysis could be done in the HS1. The vertical dot-dashed (green) line 
	        marks the transition between HS2 and HS3.}
\end{figure*}

\textit{(\romannumeral 3) Second Harder State (HS2):} 
After 2012 July 24 (MJD 56132) the soft fluxes decreased while the hard fluxes increased, implying the source has transitioned 
into a second harder state (HS2).
During this phase of the outburst, both HRs increased initially and then became constant at higher values. HR1 and HR2 varied in the 
ranges $\sim0.22-1.18$ and $\sim 0.24-0.69$ respectively. 
The $\dot{m}_h$ started to increase steadily from the SS1 to HS2 transition day and reached its maximum value on 2012 August 8 (MJD 56143). 
After that, it started to decrease. 
The $\dot{m}_d$ was observed to decrease steadily from the transition day (MJD 56132). We see an outgoing shock (as $X_s$ increased 
from $\sim 46~r_s$ to $\sim 248~r_s$) with an increasing shock strength (as $R$ increased from $\sim 1.1$ to $\sim 2.25$) during this 
phase of the outburst. We defined 2021 August 12 (MJD 56149) as the HS2 to HS3 transition day (online green dot-dashed vertical line 
in Figs. 1, 4, \& 5). 
Although we defined MJD 56132-56149 as the HS2, the peak of $\dot{m}_h$ on MJD 56143 indicates 
a possible transition from the declining hard-intermediate to the hard state.

\textit{(\romannumeral 4) Third Harder State (HS3):} From the HS2 to HS3 transition day (2012 August 12; MJD 56149),
we see a decrease in both the accretion rates. In the later phase of this state, we see an abrupt rise in the 
halo rate which attained a local maximum on 2012 September 3 (MJD 56173). This could possibly indicate the transition from rising hard to 
the hard-intermediate state. 
During this HS3 phase of the outburst, shock became weaker (changed 
from $2.25$ to $1.08$) and moved steadily inward from 248 $r_s$ to 32 $r_s$. Three type-C QPOs were also observed in three observations 
with frequencies between 2.60 Hz and 2.79 Hz. We defined 2012 September 11 (MJD 56181) as HS3 to SS2 transition day. On this day 
$\dot{m}_h$ dropped to a very low value and ARR also decreased to a very low value.

\begin{figure*}
\vskip 1.5cm
  \centering
    \includegraphics[angle=0,width=14cm,keepaspectratio=true]{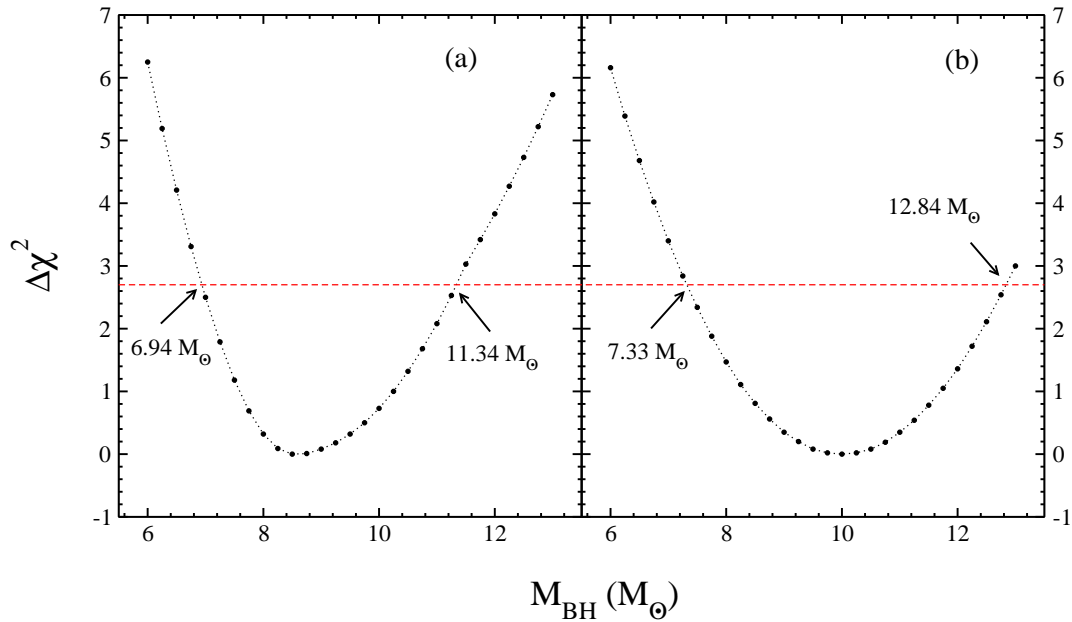}
	\caption{Variations of $\Delta\chi^2$ with $M_{BH}$ parameter variations for ObsIDs (a) 00032480009 and (b) 00032521036 taken from SS1 and HS4. The red dashed line denotes 90\% confidence level ($\Delta\chi^2=2.71$).} 
\end{figure*}

\textit{(\romannumeral 5) Second Softer State (SS2):} After 2012 September 11 (MJD 56181), the soft fluxes started to rise while 
the hard fluxes continued decreasing. As a result, the hardness ratios rapidly decreased (HR1 $\sim0.09$, HR2 $\sim0.19$).
The disk rate started to rise and the halo rate decreased to a very low value. 
We observed a clear dominance of the $\dot{m}_d$ over the $\dot{m}_h$ during this phase of the outburst. 
The ARR decreased to a level comparable to the values in the previous softer state. Presence of a weaker shock close to the black hole 
was observed. The HRs decreased to typical softer state values (HR1 varied between $\sim 0.01-0.51$ \& HR2 varied between $\sim0.04-0.33$) 
owing to an increase in the softer X-ray fluxes and a decrease in the harder X-ray fluxes. We defined 2012 October 11 (MJD 56211) as the 
transition day from this SS2 to HS4.

\textit{(\romannumeral 6) Fourth Harder State (HS4):} Starting from the transition day (SS2 to HS4; MJD 56211), we observed an increase in the halo rate 
and a decrease in the disk rate. As the outburst progressed in this declining harder state, the shock was found to become stronger ($R$ evolved from 1.1 to 3.66) 
and moved outward ($X_s$ evolved from $43$~$r_s$ to $69$~$r_s$). 
During this phase of the outburst, HRs were observed to be in a higher range as the hard fluxes dominated the soft flux. 
Two type-C QPOs were found in this state on MJD 56227 and MJD 56243 with frequencies 1.26 Hz and 1.52 Hz respectively.

\subsubsection{Estimation of BH mass from spectral analysis}

During spectral fitting with the TCAF model, one may freeze the BH mass parameter ($M_{BH}$) to a predetermined value, or it might be 
kept free to obtain the best-fitted value of mass from each spectral fit. As the exact mass of MAXI J1910-057 is not determined 
dynamically, we tried to find the most probable value of the BH mass by keeping $M_{BH}$ as free while fitting spectra 
with the TCAF model. We find the best fitted mass values to vary between 7.4--13.5 $M_\odot$ from our 35 spectral fits. This estimated 
mass range is smaller than the previously determined mass range (2.9--12.9 $M_\odot$) on the basis of inner disk radius estimation and 
luminosity considerations (Nakahira et al. 2014). It is to be noted that the variations of $M_{BH}$ in our spectral fitting (Table A) 
is a manifestation of measurement errors. As $M_{BH}$ varies as the fourth power of the inner disk temperature ($M_{BH}\sim T^4$),
small errors in the determination of the disk temperature give rise to significantly larger errors in mass determination.
We also assume that the inclination angle of the disk does not change with time, in which case the normalization of the TCAF model
would vary greatly, and this would introduce additional errors in the mass determination. Assuming the errors to be random, 
we take an average of all the best fitted $M_{BH}$ values which comes out to be $\sim9.97$~$M_\odot$.

We also employ a different method to estimate the lower and upper limits of the mass value as was done in previous papers by 
our group (Chatterjee et al. 2016; Molla et al. 2017). After obtaining the best model fit, we refit all spectra 
by keeping all other parameters frozen, except $M_{BH}$. Then $M_{BH}$ was varied manually in both +ve and -ve sides of the best fitted 
$M_{BH}$ and $\Delta\chi^2$ values were noted. The fits got deteriorated as the mass values moved away from the best fitted $M_{BH}$. 
The variation of $\Delta\chi^2$ with $M_{BH}$ is shown for two observations in Figure 7 for two sample spectra. Within acceptable i.e., 
90\% confidence interval ($\Delta\chi^2 \leq 2.71$), $M_{BH}$ values were found to vary between 6.31 $M_\odot$ to 13.65 $M_\odot$. 
Now, we can say that the probable mass of MAXI~J1910-057 lies within the range 6.31 $M_\odot$ to 13.65 $M_\odot$ with an average value of
9.97 $M_\odot$.

\subsubsection{Estimation of BH distance}

We estimated the distance of the black hole candidate (BHC) from our determined mass value and the XRT flux 
obtained from our spectral fit. This method has been used to determine the distances of BHCs by many studies (Homan et al. 2006, 
Miller-Jones et al. 2012, Saikia et al. 2022).
It is observed that during the transition from a softer state to a harder state, the luminosity of a BH binary source stays between 
0.3\% to 3\% of Eddington luminosity ($L_{Edd}$; Kalemci et al. 2013). 
From our estimated mass of 9.97 $M_\odot$, we calculated the Eddington luminosity of the source to be $\sim$ $1.3\times10^{39}$ erg s$^{-1}$. 
For a source distance of $D$ cm, we calculated the spectral state transition luminosity ($L_t$) to be $\sim$ $115.24\times10^{-9}\times D^2$ erg s$^{-1}$
from $0.1$ to $500$ keV bolometric flux on MJD 56132.23 when the source made a transition from SS1 to HS2. 
Evoking the above-mentioned constraints, we estimate the source distance to be between 1.9--5.9 kpc. 
Similarly, we also estimate the distance using SS2 to HS4 transition luminosity on MJD 56211.32. 
The spectral state transition luminosity ($L_t$) was $\sim$ $58.61\times10^{-9}\times D^2$ erg s$^{-1}$ for a source distance of $D$ cm.
The estimated distance for 0.3\% and 3\% of Eddington luminosity are 2.6 and 8.3 kpc respectively. Combining these 
two estimated ranges obtained from two state transitions of the source during its 2012-13 activity, we infer the distance of the BH binary as 
1.9--8.3 kpc.

\section{Discussion and Concluding Remarks}

We study the evolution of spectral and temporal properties of MAXI J1910-057 during its first observed outburst which started on 
2012 May 31 (MJD 56078). The BHC is studied for $\sim 5$ months using observational data from Swift/XRT, Swift/BAT, and MAXI/GSC 
instruments. Spectral analysis is done using the physical TCAF model on 35 observations spread over the outburst period. 
We evaluate physical flow parameters such as the Keplerian disk accretion rate ($\dot{m}_d$), the sub-Keplerian halo accretion 
rate ($\dot{m}_h$), the shock location ($X_s$), and the shock compression ratio (R) from the TCAF model fits. Evolution of these 
parameters helped us to get a more clear picture of the accretion dynamics of MAXI~J1910-057 during the outburst. 

MAXI J1910-057 showed a rare sequence of spectral evolution during its 2012-13 outburst. The source intensity in the soft X-ray band 
rose again even after entering the declining hard state of the main outburst. Only a few sources have shown somewhat similar kind 
of outbursts, e.g., MAXI J1535-571 (Sreehari et al. 2019, Parikh et al. 2019). According to the rebrightening classification method 
proposed by Zhang et al. (2019), such behavior can be classified as an outburst with a glitch, as the slope of the lightcurve after the 
rebrightening is similar to the slope of the lightcurve before it.  
The spectral states during the 2012-13 outburst of MAXI J1910-057 were identified primarily from the evolution of the spectral parameters of the TCAF model fit and from the nature of evolution 
of the soft and the hard X-ray fluxes and their ratios (HRs). Being unable to specify transition dates between hard state and hard intermediate state or soft state and soft intermediate state
from the variations of the spectral and temporal properties, here we define them as harder (hard and hard intermediate) and 
softer (soft and soft intermediate) spectral states.

\begin{figure}
%\vskip -1.4cm
  \centering
    \includegraphics[angle=0,width=\linewidth,keepaspectratio=true]{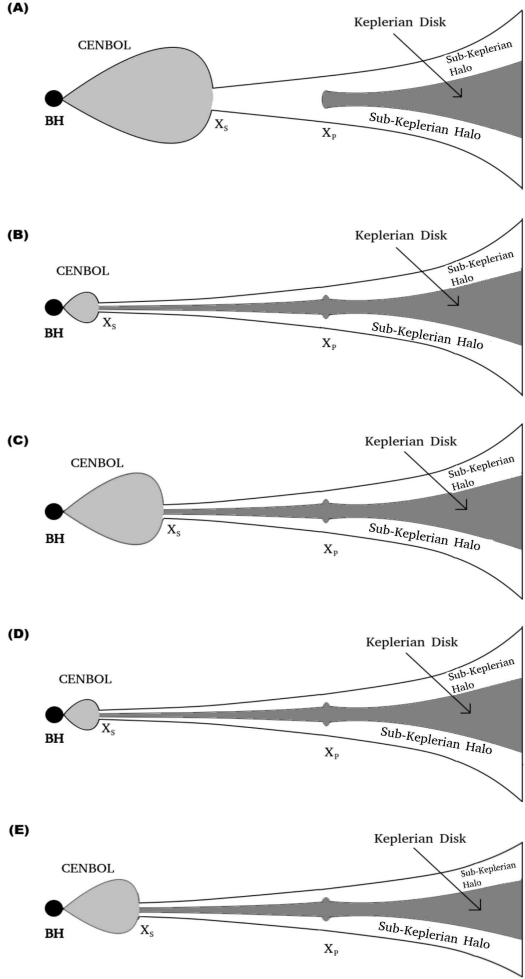}
	\caption{A diagram of the evolution of accretion flow geometry for the 2012 outburst of MAXI J1910-057. 
	(A) Just prior to the outburst Keplerian matter gets piled up at the pile-up radius ($X_P$) and the location of the shock ($X_S$) is large. 
	(B) Viscosity at the pile-up radius increases, outburst starts, and the Keplerian matter reaches towards the BH cooling and shrinking down the CENBOL (SS1).
	(C) The increased supply of sub-Keplerian matter increases the CENBOL size in HS2. In HS3, eventually the increased supply of Keplerian matter reaches the BH and the CENBOL starts to shrink. 
	(D) In SS2 the CENBOL shrinks down and the shock location becomes minimum. (E) In HS4, the supply of accreting matter starts to deplete and the outburst goes to the declining phase.} 
\end{figure}

The low rate of the halo accretion with respect to disk accretion and the presence of a weak shock close to the BH indicates 
that the accreting Keplerian matter had cooled down the CENBOL to a small size and as a result, the spectrum became soft and the state became SS1. 
As the CENBOL is the repository of hot electrons which are responsible for the inverse-Comptonization of thermal photons, only a small amount 
of the soft photons from the Keplerian disk interacted with the smaller-sized CENBOL in SS1 and became hard via inverse Comptonization. 
Overall spectra were dominated by the soft photons emitted from the Keplerian disk.
After that, the supply of the sub-Keplerian matter increased as the source moved into the declining phase of the main outburst. In this phase, 
the CENBOL region became larger, and the shock location, which denotes the outer boundary of CENBOL, began to recede away from the BH 
and increased in strength. Increase in CENBOL size resulted in an increase in the high energy X-ray flux and the spectra became hard and 
the spectral state became HS2. 
The shock reached the farthest location on 2021 August 12 (MJD 56149), and after that, the shock started to move inward with decreasing 
strength. This is quite uncommon during the evolution of a transient BHC. Around this day, we observe roughly periodic variability in HR1. 
A flux dip was also observed across all wavebands (NIR, Optical, UV, X-ray) in the following period ($\sim$ MJD 56162-56182) as 
reported by Degenaar et al. (2014). All these factors signify a change in the accretion flow morphology, possibly due to a sudden 
increase in the supply of matter from the companion or from the pile-up radius or a small-scale mass transfer instability.

From our recent findings on the nature of outbursts, accreting matter from the companion gets accumulated at the pile-up 
radius until sufficient viscosity is reached for the matter to proceed towards the BH causing an outburst (Chakrabarti et al. 2019; 
Bhowmick et al. 2021; Chatterjee et al. 2022). In general, during a declining hard state, we observe a decrease in both accretion rates, 
but the Keplerian rate decreases faster. So, we observe a dominance of the halo rate and the presence of hard spectra. But, during the 2012-13 
outburst of MAXI J1910-057, the rate of matter supply from the companion increased suddenly at the start of HS2. This might be due to a sudden 
rise in viscosity at the pile-up radius, which triggered a reflare. The fresh supply of sub-Keplerian matter increased the strength and 
size of the CENBOL. So, the shock was observed to move away. Then the shock was observed to move inward after 2021 August 12 (MJD 56149) 
as the halo rate decreased rapidly. We defined this phase of the outburst as HS3 which is very similar to the initial harder state 
of an outburst.
Then the source moved to the SS2. At the start of the SS2, the halo rate, shock location, and its strength decreased to their lowest 
values, however disk rate started to increase. After that, in the HS4, the disk rate decreased whereas the halo rate increased a little bit. 
Due to this, the cooling process became inefficient and the CENBOL began to increase in size and the shock began to move away from the BH, 
which is generally observed in the declining hard state of a BH outburst. Cartoon diagrams of accretion flow geometry during different phases of the 2012 outburst of MAXI J1910-057 are shown in Fig. 8.

We detected five low-frequency QPOs (LFQPOs), three in the HS3 with frequencies $\sim2.6 - 2.8$ Hz and the remaining two 
in the HS4 with frequencies $1.3$ and $1.5$ Hz (see, Table 1). According to the criteria mentioned in Casella et al. (2005) 
we categorized these QPOs to be of type-C based on their centroid frequencies, Q-values, rms variations, etc. These LFQPOs are 
believed to be originated from the oscillations of the shock, either due to a resonance that sets in when the cooling time scale 
of the flow becomes comparable to the infall time scale (Molteni et al. 1996) or due to the non-satisfaction of Rankine-Hugoniot 
conditions that are necessary to form stationary shocks (Ryu et al. 1997). It is to be noted that QPOs are also found at 
$\sim4.5$ Hz and $\sim6$ Hz in an XMM-NEWTON observation (Reis et al. 2013), however, we do not find any of such QPOs from 
our Swift/XRT observations.

So far, no dynamic measurements of mass have been done for this source. By determining the inner disk radius from their spectral 
fits, Nakahira et al. (2014) estimated the mass of this BH to be in the range 2.9--12.9 $M_\odot$ which is rather large. From the 
TCAF model spectral fits, we try to constrain the probable mass value of the BH in a narrower range. In the physical TCAF model,
many of its essential features responsible for producing the spectrum, such as the size and the electron number density inside 
the CENBOL, intensity of the low energy radiation from the Keplerian disk, etc. depend directly on the mass of the BH which is an 
important input parameter of this model. During spectral analysis, 
we allow the $M_{BH}$ parameter to vary freely and find the best-fitted mass values to lie in the range 7.4-13.5 $M_\odot$.
We checked the variation in the goodness of the spectral fits with different $M_{BH}$ values keeping all other parameters 
frozen. With 90\% confidence ($\Delta\chi^2 \leq 2.71$) the lower and upper limits of $M_{BH}$ values are found to be 6.31 $M_\odot$ 
and 13.65 $M_\odot$ respectively. Hence we can say the mass of the BH MAXI J1910-057 lies in the range 
6.31--13.65 $M_\odot$ with an average value of 9.97 $M_\odot$.

It has been observed that the transition luminosity for soft to hard state transition lies between 0.3\% to 3\% of Eddington 
luminosity of the source (Kalemci et al. 2013). 
We calculated the Eddington luminosity using our estimated mass of the BH. From our 
spectral fitting, we calculated the bolometric flux in 0.1-500 keV energy range. 
In the 2012 outburst, the source transits into the harder state twice in the declining phase.
Considering the transition luminosity to remain in the 0.3\% to 3\% $L_{Edd}$ limit, we estimated the source distance 
as $1.9-5.9$ kpc and $2.6-8.3$ kpc from the first and second state transitions, respectively. Combining both estimates, we infer 
the distance of MAXI~J1910-057 as $1.9-8.3$~kpc.

\section*{Acknowledgements}
We are thankful to anonymous referees for their comments and suggestions to improve the quality of the manuscript.
This work made use of \href{https://swift.gsfc.nasa.gov/results/transients/}{Swift}/XRT and /BAT data supplied by the UK Swift Science Data Centre 
at the University of Leicester, and \href{http://maxi.riken.jp/top/lc.html}{MAXI}/GSC data were provided by RIKEN, JAXA, and the MAXI team. 
These are publicly available and can be downloaded from HEASARC's \href{https://heasarc.gsfc.nasa.gov/cgi-bin/W3Browse/w3browse.pl}{W3Browse} 
or satellite websites. The current version of the TCAF model {\it fits} file which is used in this paper, is now publicly available on-demand basis. 
Contact the corresponding author of this paper for further details.
Authors S.K.N. and D.D. acknowledge support from ISRO sponsored RESPOND project (ISRO/RES/2/418/17-18) fund.
K.C. acknowledges support from DST/INSPIRE Fellowship (IF170233). 
Research of D.D. is supported in part by the Higher Education Dept. 
of the Govt. of West Bengal, India. 
D.D. also acknowledges support from DST/GITA sponsored India–Taiwan collaborative project (GITA/DST/TWN/P-76/2017) fund. 
A.J. acknowledges post-doctoral fellowship of Physical Research Laboratory, Ahmedabad, funded by the Department of Space, Government of India. 
R.B. acknowledges support from CSIR-UGC fellowship (June-2018, 527223).

\clearpage

\vskip -1cm
\begin{table*}
 \addtolength{\tabcolsep}{-2.0pt}
 \centering
 \caption{Best fitted spectral model parameters}
 \label{tab:table2}
                \begin{tabular}{ccccc|ccccccc}
\hline
    Obs Id  &  UT$^a$  & Day     & XRT     & BAT     & $\dot{m}_d$$^b$   & $\dot{m}_h$$^b$    & ARR$^b$ & $X_s$$^b$ & $R$$^b$  & $\chi^2/dof$$^c$ \\
	    & Date     & (MJD)   & exp (s) & exp (s) & ($\dot{M}_{Edd}$) & ($\dot{M}_{Edd}$)  &         & $(r_s)$   &          &                     \\
  (1)       &   (2)    & (3)     &   (4)   &   (5)   &   (6)             &    (7)             &    (8)  &     (9)   &  (10)    &   (11)          \\
\hline\hline
00032480002 & 06-04 & 56082.5 &  999.4 &  ...   & $0.46^{\pm0.01}$ & $0.01^{\pm0.01}$ & $0.03^{\pm0.01}$ & $ 56^{\pm5}$ & $1.08^{\pm0.02}$  & $ 516/349 $ \\
00032480005 & 06-14 & 56092.2 & 1483.9 &  ...   & $0.98^{\pm0.01}$ & $0.02^{\pm0.01}$ & $0.02^{\pm0.01}$ & $ 55^{\pm3}$ & $1.08^{\pm0.12}$  & $ 555/383 $ \\
00524642000 & 06-18 & 56096.8 & 2525.9 &  ...   & $0.84^{\pm0.01}$ & $0.07^{\pm0.01}$ & $0.09^{\pm0.01}$ & $ 56^{\pm4}$ & $1.09^{\pm0.03}$  & $ 618/449 $ \\
00032480007 & 06-20 & 56098.8 & 1119.8 &  ...   & $0.82^{\pm0.01}$ & $0.07^{\pm0.01}$ & $0.09^{\pm0.01}$ & $ 56^{\pm4}$ & $1.09^{\pm0.04}$  & $ 615/483 $ \\
00032480008 & 06-23 & 56101.1 & 1169.2 &  ...   & $0.78^{\pm0.01}$ & $0.07^{\pm0.02}$ & $0.09^{\pm0.01}$ & $ 56^{\pm5}$ & $1.08^{\pm0.01}$  & $ 489/408 $ \\
00032480009 & 06-26 & 56104.5 &  939.6 &  ...   & $0.52^{\pm0.01}$ & $0.07^{\pm0.01}$ & $0.14^{\pm0.01}$ & $ 50^{\pm6}$ & $1.05^{\pm0.03}$  & $ 446/444 $ \\
00032480010 & 06-29 & 56107.5 & 1114.6 &  ...   & $0.24^{\pm0.01}$ & $0.02^{\pm0.01}$ & $0.09^{\pm0.02}$ & $ 41^{\pm7}$ & $1.05^{\pm0.00}$  & $ 466/353 $ \\
00032480011 & 07-02 & 56110.3 &  979.6 &  ...   & $0.24^{\pm0.01}$ & $0.02^{\pm0.01}$ & $0.09^{\pm0.01}$ & $ 41^{\pm7}$ & $1.05^{\pm0.01}$  & $ 453/336 $ \\
00032480014 & 07-17 & 56125.4 &  999.4 &  ...   & $0.16^{\pm0.01}$ & $0.01^{\pm0.01}$ & $0.03^{\pm0.01}$ & $ 39^{\pm2}$ & $1.08^{\pm0.00}$  & $ 412/289 $ \\
\hline
00032480015 & 07-24 & 56132.2 &  763.7 &  ...   & $0.16^{\pm0.03}$ & $0.03^{\pm0.01}$ & $0.21^{\pm0.05}$ & $ 47^{\pm2}$ & $1.07^{\pm0.01}$  & $ 355/318 $ \\
00529076000 & 07-29 & 56137.3 & 3154.7 & 6881.4 & $0.15^{\pm0.02}$ & $0.09^{\pm0.01}$ & $0.59^{\pm0.08}$ & $175^{\pm4}$ & $1.08^{\pm0.01}$  & $1905/560 $ \\
00529076001 & 07-29 & 56137.4 &  158.9 &  ...   & $0.15^{\pm0.02}$ & $0.08^{\pm0.02}$ & $0.55^{\pm0.11}$ & $172^{\pm2}$ & $1.07^{\pm0.04}$  & $ 184/162 $ \\
00032521001 & 08-04 & 56143.8 &  984.4 &  ...   & $0.11^{\pm0.03}$ & $0.12^{\pm0.01}$ & $1.09^{\pm0.30}$ & $215^{\pm3}$ & $1.31^{\pm0.08}$  & $ 266/220 $ \\
00032521002 & 08-07 & 56146.5 &  979.6 &  ...   & $0.10^{\pm0.02}$ & $0.11^{\pm0.02}$ & $1.15^{\pm0.23}$ & $227^{\pm2}$ & $1.68^{\pm0.12}$  & $ 314/268 $ \\
\hline\hline
00032521003 & 08-10 & 56149.4 & 1204.2 &  ...   & $0.09^{\pm0.01}$ & $0.11^{\pm0.01}$ & $1.17^{\pm0.11}$ & $248^{\pm6}$ & $2.25^{\pm0.02}$  & $ 759/519 $ \\
00032521004 & 08-13 & 56152.4 & 1074.6 &  ...   & $0.08^{\pm0.01}$ & $0.09^{\pm0.01}$ & $1.12^{\pm0.22}$ & $238^{\pm3}$ & $2.19^{\pm0.05}$  & $ 528/440 $ \\
00032521005 & 08-16 & 56155.4 & 1074.4 &  ...   & $0.08^{\pm0.04}$ & $0.08^{\pm0.01}$ & $1.03^{\pm0.50}$ & $238^{\pm4}$ & $2.13^{\pm0.03}$  & $ 587/380 $ \\
00032521006 & 08-19 & 56158.7 &  558.1 &  ...   & $0.06^{\pm0.01}$ & $0.07^{\pm0.02}$ & $1.09^{\pm0.25}$ & $229^{\pm8}$ & $2.08^{\pm0.02}$  & $ 479/348 $ \\
00032521007 & 08-22 & 56161.0 & 1049.6 &  ...   & $0.07^{\pm0.01}$ & $0.08^{\pm0.01}$ & $1.07^{\pm0.06}$ & $215^{\pm4}$ & $1.98^{\pm0.04}$  & $ 673/438 $ \\
00032521008 & 08-25 & 56164.7 &  924.4 &  ...   & $0.06^{\pm0.01}$ & $0.05^{\pm0.01}$ & $0.95^{\pm0.11}$ & $176^{\pm5}$ & $1.84^{\pm0.02}$  & $ 586/396 $ \\
00032521009 & 08-29 & 56168.8 & 1074.5 &  ...   & $0.05^{\pm0.01}$ & $0.04^{\pm0.01}$ & $0.92^{\pm0.25}$ & $144^{\pm3}$ & $1.65^{\pm0.01}$  & $ 567/369 $ \\
00032521010 & 09-01 & 56171.3 &  964.2 &  ...   & $0.03^{\pm0.01}$ & $0.04^{\pm0.01}$ & $1.25^{\pm0.39}$ & $ 85^{\pm1}$ & $1.81^{\pm0.01}$  & $ 457/286 $ \\
00032521011 & 09-03 & 56173.4 &  979.6 &  ...   & $0.04^{\pm0.01}$ & $0.06^{\pm0.01}$ & $1.61^{\pm0.28}$ & $ 70^{\pm2}$ & $1.83^{\pm0.01}$  & $ 338/231 $ \\
00032521012 & 09-05 & 56175.4 & 1014.6 &  ...   & $0.02^{\pm0.01}$ & $0.02^{\pm0.01}$ & $0.89^{\pm0.38}$ & $ 41^{\pm5}$ & $1.76^{\pm0.12}$  & $ 300/203 $ \\
00032521013 & 09-07 & 56177.8 &  969.4 &  ...   & $0.01^{\pm0.01}$ & $0.01^{\pm0.01}$ & $0.43^{\pm0.12}$ & $ 40^{\pm2}$ & $1.67^{\pm0.08}$  & $ 365/244 $ \\
\hline
00032521015 & 09-11 & 56181.5 &  534.4 &  ...   & $0.02^{\pm0.01}$ & $0.01^{\pm0.01}$ & $0.06^{\pm0.04}$ & $ 32^{\pm0}$ & $1.08^{\pm0.03}$  & $ 273/192 $ \\
00032521017 & 09-15 & 56185.0 &  948.2 &  ...   & $0.05^{\pm0.01}$ & $0.01^{\pm0.01}$ & $0.05^{\pm0.01}$ & $ 33^{\pm6}$ & $1.08^{\pm0.07}$  & $ 350/268 $ \\
00032521018 & 09-17 & 56187.7 & 1009.6 &  ...   & $0.07^{\pm0.02}$ & $0.01^{\pm0.01}$ & $0.10^{\pm0.03}$ & $ 38^{\pm3}$ & $1.05^{\pm0.01}$  & $ 382/260 $ \\
00032521021 & 09-25 & 56195.8 &  974.6 &  ...   & $0.05^{\pm0.01}$ & $0.01^{\pm0.01}$ & $0.06^{\pm0.11}$ & $ 33^{\pm1}$ & $1.09^{\pm0.07}$  & $ 437/283 $ \\
00032521025 & 10-03 & 56203.4 &  999.4 &  ...   & $0.07^{\pm0.04}$ & $0.01^{\pm0.01}$ & $0.02^{\pm0.02}$ & $ 39^{\pm4}$ & $1.08^{\pm0.11}$  & $ 365/263 $ \\
\hline
00032521029 & 10-11 & 56211.3 &  964.6 &  ...   & $0.07^{\pm0.01}$ & $0.01^{\pm0.01}$ & $0.16^{\pm0.07}$ & $ 43^{\pm1}$ & $1.10^{\pm0.02}$  & $ 507/351 $ \\
00032521032 & 10-17 & 56217.5 & 1054.6 &  ...   & $0.06^{\pm0.01}$ & $0.04^{\pm0.01}$ & $0.63^{\pm0.08}$ & $ 51^{\pm5}$ & $1.81^{\pm0.03}$  & $ 523/351 $ \\
00032521036 & 10-27 & 56227.2 &  999.5 &  ...   & $0.03^{\pm0.01}$ & $0.06^{\pm0.01}$ & $2.14^{\pm0.28}$ & $ 54^{\pm4}$ & $2.49^{\pm0.06}$  & $ 420/301 $ \\
00032521040 & 11-04 & 56235.0 &  959.6 &  ...   & $0.01^{\pm0.01}$ & $0.05^{\pm0.01}$ & $3.43^{\pm1.02}$ & $ 69^{\pm1}$ & $3.41^{\pm0.09}$  & $ 352/232 $ \\
00032521045 & 11-14 & 56245.8 & 1099.3 &  ...   & $0.01^{\pm0.01}$ & $0.04^{\pm0.01}$ & $3.73^{\pm1.26}$ & $ 69^{\pm2}$ & $3.66^{\pm0.14}$  & $ 274/198 $ \\
\hline
                \end{tabular}
\vskip 0.2cm
\noindent{
	\leftline{$^a$ UT Date in mm-dd format and year of observation is 2012.} 
        \leftline{$^b$ TCAF model fitted spectral parameters with fixed $M_{BH}$ ($=9.97~M_\odot$) are shown in columns (6)-(10). All parameter}
        \leftline{ and errors except $X_s$ have been approximated up to the second decimal place.} 
        \leftline{$^c$ $\chi^2$ values and degrees of freedom (dof) of the TCAF spectral model fits are shown in columns (11).}
        \leftline{\textit{Note.} Superscripts on the parameter values represent average error values obtained using the {\fontfamily{qcr}\selectfont err} task in {\fontfamily{qcr}\selectfont XSPEC} with} 
        \leftline{90\% confidence.}
}

\end{table*}

\begin{table*}
 \addtolength{\tabcolsep}{-2.0pt}
 \centering
 \centering{\Large Appendix}
 \vskip 0.1cm
 \centering{ Table A: Preliminary spectral analysis parameters with $M_{BH}$ as a free parameter}
\vskip 0.2cm
                \begin{tabular}{cc|cccccc}
\hline
 Obs Id  &    Day  & $\dot{m}_d$$^a$   & $\dot{m}_h$$^a$    &  $M_{BH}$$^a$ & $X_s$$^a$ & $R$$^a$  & $\chi^2/dof$$^b$ \\
         & (MJD)   & ($\dot{M}_{Edd}$) & ($\dot{M}_{Edd}$)  &  ($M_\odot$)  & $(r_s)$   &          &                  \\
  (1)    &   (2)   &       (3)         &        (4)         &      (5)      &    (6)    &   (7)    &       (8)        \\
\hline\hline
 00032480002 & 56082.5 & $0.45^{\pm0.02}$ & $0.01^{\pm0.01}$ & $10.0^{\pm0.1}$ & $ 56^{\pm7}$ & $1.09^{\pm0.03}$ & $ 513/348 $ \\
 00032480005 & 56092.2 & $0.97^{\pm0.01}$ & $0.02^{\pm0.01}$ & $11.4^{\pm0.6}$ & $ 55^{\pm2}$ & $1.08^{\pm0.12}$ & $ 552/382 $ \\
 00524642000 & 56096.8 & $0.84^{\pm0.03}$ & $0.07^{\pm0.01}$ & $ 9.0^{\pm0.2}$ & $ 56^{\pm4}$ & $1.10^{\pm0.04}$ & $ 615/448 $ \\
 00032480007 & 56098.8 & $0.82^{\pm0.02}$ & $0.07^{\pm0.02}$ & $ 9.7^{\pm0.3}$ & $ 56^{\pm3}$ & $1.10^{\pm0.05}$ & $ 604/482 $ \\
 00032480008 & 56101.1 & $0.77^{\pm0.01}$ & $0.07^{\pm0.02}$ & $ 9.6^{\pm0.2}$ & $ 55^{\pm5}$ & $1.09^{\pm0.02}$ & $ 495/407 $ \\
 00032480009 & 56104.5 & $0.51^{\pm0.01}$ & $0.07^{\pm0.01}$ & $ 8.5^{\pm0.2}$ & $ 51^{\pm9}$ & $1.05^{\pm0.03}$ & $ 441/443 $ \\
 00032480010 & 56107.5 & $0.24^{\pm0.01}$ & $0.02^{\pm0.01}$ & $ 8.5^{\pm0.2}$ & $ 42^{\pm6}$ & $1.05^{\pm0.03}$ & $ 452/352 $ \\
 00032480011 & 56110.3 & $0.24^{\pm0.01}$ & $0.02^{\pm0.01}$ & $ 8.6^{\pm0.1}$ & $ 43^{\pm8}$ & $1.05^{\pm0.01}$ & $ 435/335 $ \\
 00032480014 & 56125.4 & $0.16^{\pm0.01}$ & $0.01^{\pm0.01}$ & $12.9^{\pm0.6}$ & $ 39^{\pm5}$ & $1.09^{\pm0.05}$ & $ 399/288 $ \\
\hline
 00032480015 & 56132.2 & $0.16^{\pm0.02}$ & $0.03^{\pm0.01}$ & $12.9^{\pm0.3}$ & $ 44^{\pm2}$ & $1.06^{\pm0.02}$ & $ 419/317 $ \\
 00529076000 & 56137.3 & $0.15^{\pm0.01}$ & $0.09^{\pm0.01}$ & $ 7.8^{\pm0.0}$ & $176^{\pm4}$ & $1.09^{\pm0.02}$ & $1857/559 $ \\
 00529076001 & 56137.4 & $0.15^{\pm0.02}$ & $0.08^{\pm0.02}$ & $ 8.1^{\pm0.5}$ & $173^{\pm2}$ & $1.07^{\pm0.05}$ & $ 184/161 $ \\
 00032521001 & 56143.8 & $0.11^{\pm0.03}$ & $0.12^{\pm0.01}$ & $ 8.4^{\pm0.2}$ & $216^{\pm4}$ & $1.31^{\pm0.19}$ & $ 266/219 $ \\
 00032521002 & 56146.5 & $0.10^{\pm0.02}$ & $0.11^{\pm0.02}$ & $ 8.4^{\pm0.4}$ & $211^{\pm3}$ & $1.68^{\pm0.14}$ & $ 314/267 $ \\
\hline\hline 
 00032521003 & 56149.4 & $0.09^{\pm0.01}$ & $0.11^{\pm0.01}$ & $ 8.4^{\pm0.1}$ & $248^{\pm7}$ & $2.25^{\pm0.04}$ & $ 759/518 $ \\
 00032521004 & 56152.4 & $0.08^{\pm0.02}$ & $0.09^{\pm0.02}$ & $ 7.6^{\pm0.1}$ & $238^{\pm5}$ & $2.19^{\pm0.05}$ & $ 528/439 $ \\
 00032521005 & 56155.4 & $0.08^{\pm0.04}$ & $0.08^{\pm0.01}$ & $ 7.7^{\pm0.1}$ & $238^{\pm3}$ & $2.13^{\pm0.03}$ & $ 579/379 $ \\
 00032521006 & 56158.7 & $0.06^{\pm0.01}$ & $0.07^{\pm0.02}$ & $12.6^{\pm0.1}$ & $229^{\pm8}$ & $2.08^{\pm0.05}$ & $ 479/347 $ \\
 00032521007 & 56161.0 & $0.07^{\pm0.01}$ & $0.08^{\pm0.01}$ & $12.8^{\pm0.2}$ & $215^{\pm4}$ & $1.98^{\pm0.04}$ & $ 673/437 $ \\
 00032521008 & 56164.7 & $0.06^{\pm0.01}$ & $0.05^{\pm0.01}$ & $12.9^{\pm0.1}$ & $176^{\pm2}$ & $1.85^{\pm0.03}$ & $ 563/395 $ \\
 00032521009 & 56168.8 & $0.05^{\pm0.02}$ & $0.04^{\pm0.01}$ & $ 7.5^{\pm0.1}$ & $144^{\pm1}$ & $1.65^{\pm0.02}$ & $ 541/368 $ \\
 00032521010 & 56171.3 & $0.03^{\pm0.01}$ & $0.04^{\pm0.01}$ & $ 7.5^{\pm0.1}$ & $ 85^{\pm1}$ & $1.82^{\pm0.02}$ & $ 426/285 $ \\
 00032521011 & 56173.4 & $0.04^{\pm0.01}$ & $0.06^{\pm0.02}$ & $ 7.5^{\pm0.2}$ & $ 70^{\pm2}$ & $1.84^{\pm0.02}$ & $ 309/230 $ \\
 00032521012 & 56175.4 & $0.02^{\pm0.01}$ & $0.02^{\pm0.01}$ & $ 8.4^{\pm0.2}$ & $ 41^{\pm8}$ & $1.76^{\pm0.14}$ & $ 297/202 $ \\
 00032521013 & 56177.8 & $0.02^{\pm0.01}$ & $0.01^{\pm0.01}$ & $10.6^{\pm0.5}$ & $ 40^{\pm2}$ & $1.67^{\pm0.13}$ & $ 356/243 $ \\
\hline
 00032521015 & 56181.5 & $0.02^{\pm0.01}$ & $0.01^{\pm0.01}$ & $11.5^{\pm0.3}$ & $ 31^{\pm1}$ & $1.08^{\pm0.05}$ & $ 259/191 $ \\
 00032521017 & 56185.0 & $0.05^{\pm0.01}$ & $0.01^{\pm0.01}$ & $11.6^{\pm0.4}$ & $ 32^{\pm2}$ & $1.08^{\pm0.08}$ & $ 334/267 $ \\
 00032521018 & 56187.7 & $0.08^{\pm0.03}$ & $0.02^{\pm0.01}$ & $ 8.4^{\pm0.6}$ & $ 39^{\pm9}$ & $1.05^{\pm0.03}$ & $ 367/259 $ \\
 00032521021 & 56195.8 & $0.05^{\pm0.02}$ & $0.01^{\pm0.02}$ & $11.4^{\pm0.8}$ & $ 33^{\pm1}$ & $1.09^{\pm0.07}$ & $ 430/282 $ \\
 00032521025 & 56203.4 & $0.07^{\pm0.04}$ & $0.01^{\pm0.01}$ & $11.0^{\pm0.4}$ & $ 38^{\pm5}$ & $1.09^{\pm0.11}$ & $ 358/262 $ \\
\hline
 00032521029 & 56211.3 & $0.07^{\pm0.02}$ & $0.01^{\pm0.01}$ & $11.5^{\pm0.2}$ & $ 43^{\pm1}$ & $1.10^{\pm0.02}$ & $ 494/350 $ \\
 00032521032 & 56217.5 & $0.0r^{\pm0.02}$ & $0.04^{\pm0.01}$ & $11.2^{\pm0.5}$ & $ 51^{\pm6}$ & $1.80^{\pm0.04}$ & $ 522/350 $ \\
 00032521036 & 56227.2 & $0.03^{\pm0.01}$ & $0.06^{\pm0.02}$ & $ 9.9^{\pm0.1}$ & $ 54^{\pm5}$ & $2.55^{\pm0.09}$ & $ 415/300 $ \\
 00032521040 & 56235.0 & $0.02^{\pm0.01}$ & $0.05^{\pm0.01}$ & $12.8^{\pm0.2}$ & $ 69^{\pm2}$ & $3.56^{\pm0.11}$ & $ 347/231 $ \\
 00032521045 & 56245.8 & $0.01^{\pm0.01}$ & $0.04^{\pm0.02}$ & $11.9^{\pm0.2}$ & $ 69^{\pm2}$ & $3.66^{\pm0.17}$ & $ 269/197 $ \\
\hline     
                \end{tabular}
\vskip 0.2cm
\noindent{
        \leftline{$^a$ TCAF model fitted spectral parameters are shown in columns (3)-(7). Here, all parameters including $M_{BH}$}
        \leftline{are kept as free.}
        \leftline{Parameter values and errors except $M_{BH}$ and $X_s$ have been approximated up to the second decimal place.}
	\leftline{$^b$ $\chi^2$ values and degrees of freedom (dof) of the TCAF spectral model fits are shown in column (8).}
        \leftline{\textit{Note.} Superscripts on the parameter values represent average error values obtained using the {\fontfamily{qcr}\selectfont err}
                           task in}
                           \leftline{{\fontfamily{qcr}\selectfont XSPEC} with 90\% confidence.}

}

\end{table*}

\end{document}